# JigSaw: A tool for discovering explanatory high-order interactions from random forests

Demetrius DiMucci[1*]
[1] Department of Microbiology, The Forsyth Institute, Cambridge, MA, United States
[*] Corresponding author
E-mail: ddimucci@forsyth.org

## Abstract

Machine learning is revolutionizing biology by facilitating the prediction of outcomes from complex patterns found in massive data sets. Large biological data sets, like those generated by transcriptome or microbiome studies, measure many relevant components that interact *in vivo* with one another in modular ways. Identifying the high-order interactions that machine learning models use to make predictions would facilitate the development of hypotheses linking combinations of measured components to outcome. By using the structure of random forests, a new algorithmic approach, termed JigSaw, was developed to aid in the discovery of patterns that could explain predictions made by the forest. By examining the patterns of individual decision trees JigSaw identifies high-order interactions between measured features that are strongly associated with a particular outcome and identifies the relevant decision thresholds. JigSaw's effectiveness was tested in simulation studies where it was able to recover multiple ground truth patterns; even in the presence of significant noise. It was then used to find patterns associated with outcomes in two real world data sets. It was first used to identify patterns clinical measurements associated with heart disease. It was then used to find patterns associated with breast cancer using metabolites measured in the blood. In heart disease, JigSaw identified several three-way interactions that combine to explain most of the heart disease records (66%) with high precision (93%). In breast cancer, three two-way interactions were recovered that can be combined to explain almost all records (92%) with good precision (79%). JigSaw is an efficient method for exploring high-dimensional feature spaces for rules that explain statistical associations with a given outcome and can inspire the generation of testable hypotheses.

## Introduction

It is now common practice to use machine learning identify biomarkers for classifying interesting biological outcomes using data sets that contain many predictive features (e.g. gene expression in cancer, microbial abundances in Crohn's disease) (1–3) . These data sets frequently contain many statistical interactions between features, which can sometimes be high-order (i.e. involve three or more features) (4). Often, the most successful algorithms on these data sets produce complicated models whose inner workings are difficult to interpret (5). When only prediction is required, it is enough to leave such a black box unexamined. However, there is an increasing need to understand why particular predictions are made, especially in health care where it is being prioritized (6,7). The ability to explain black box decisions facilitates the development of mechanistic hypotheses connecting features to outcome (8,9) and can reveal inappropriate biases that might otherwise go undetected (10).

In the life sciences, random forest has become a popular algorithm for classification tasks because it works well on data with many high-order interactions (11). Random forest is an approach that uses ensembles of binary decision trees to classify data (12). The method is useful because it is non-parametric and innately detects high-order interactions between features (13). The ensemble method, which is the cause of much of its prognostic success, also works to impede explanation (14). Recent work showing how to quantify the effect individual features have on classification probabilities has significantly contributed to making the explanation of individual predictions easier (e.g. feature contributions, Shapley values) (15–17), but these methods do not explicitly identify sets of interacting features. Other approaches search the forest for interactions by identifying frequently co-occurring sets of features either by extracting all rules found in the forest (18) or by regularization (19). These approaches can successfully identify high-order interactions when there is one or two such sets in the data. However, the presence of multiple sets of interacting features can cause these methods to produce chimeric rules, likely in part due to masking effects (20). The presence of spurious labels can further cause these methods to identify misleading explanatory rule sets.

In a previous work by the author an algorithmic approach for identifying sets of interacting categorical features called BowSaw was described (21). BowSaw uses the structure of individual trees in the forest to construct explanatory rules that are strongly associated with a particular outcome. The BowSaw algorithm works by following the paths a single specified sample takes through its out-of-bag trees of the forest. It ranks all parent-child pairwise feature co-occurrences the sample encounters by their frequency and uses these ranked interactions as a guide to greedily construct a set of features maximally associated with the observed outcome of the sample. It was demonstrated through simulations that the BowSaw method can fully recover the majority of multiple high-order sets of interacting features even in the presence of labeling noise. BowSaw is limited however, in that it can only be applied to categorical data.

In this work, JigSaw extends the capabilities of BowSaw to handle continuous feature spaces. For an individual sample, JigSaw uses the original BowSaw framework to rank pairwise feature interactions. It then builds a set of likely interacting features by greedily adding high-ranking feature pairs to the set and using the Euclidean distance in those dimensions to identify a set that maximizes local enrichment for the observed outcome of the sample. Once a likely set of features is identified, it then retraces the paths through the forest in order to find approximations for the relevant upper and lower decision boundaries. It is first demonstrated on simulated data sets that JigSaw is capable of fully recovering multiple true interacting sets, even in the presence of spurious labeling. It is also shown that the putative decision boundaries identified by JigSaw cover most of the true decision space. JigSaw is then applied to two real world case studies in order to derive explanatory patterns associated with outcomes. The first is the UCI machine learning repository heart disease dataset where measurements of 13 clinically relevant features are provided for each record (22). JigSaw identified several simple rules and accompanying decision boundaries that account for 66% of heart disease records with 92% precision. The second data set is a breast cancer data set where measurements of metabolites in the blood are provided for each record (23). In this instance, JigSaw identified a concise set of rules and boundaries that combined to explain 92% of breast cancer records with 78% precision.

## Methods

The JigSaw algorithm identifies interacting features relevant to a specified observation in two steps. First, it counts the frequency with which the observation encounters instances of parent-child feature pairs (disregarding order) along the relevant decision paths in its corresponding out-of-bag trees. Second, it uses a ranked set of feature pairs to construct a high-order feature set that maximizes the observations' local association with other observed points from the same class label. The symbols used in algorithms 1 and 2 are further defined in Table 1.

### Algorithm 1 – Counting parent-child feature pairs for a specified observation
$M$ = trained random forest model
$T$ = set out out-of-bag trees in $M$
Initialize matrix $N$ as a $p$ x $p$ matrix of zeros
**for** each tree in $T$ **do**
    *path* = sequence of feature indices in tree
    **for** index in 2: length(path) **do**
        *parent* = *path*[index – 1]
        *child* = *path*[index]
        *first* = max(*parent, child*)

     *second* = min(*parent, child*)
     *N*[*first, second*] = *N*[*first, second*] + 1
   **end for**
 **end for**
**end for**
**return** *N*

## Algorithm 2 – identify an associated feature set for a specific observation

*D* = dataset of continuous values
*Y* = Observed labels vector (1 if same as target observation, 0 otherwise)
*V* = vector of the target observations' values
*A* = vector of zeros of length #*Y*
*E* = Set of pairs derived from *N*, sorted by descending frequency
*R* = ∅
*count* = 0
**while** *count* < pairs in *E* **do**
  *count* = *count* + 1
  $R^* = \cup(R, E_{count})$
  $vals = V_{R^*}$
  $D^* = (D[1:end, R^*] - vals)^2$
  $S = \begin{bmatrix} \sqrt{\sum D^*}_1 \\ ... \\ \sqrt{\sum D^*}_2 \\ ... \\ \sqrt{\sum D^*}_{end} \end{bmatrix}$
  *k* = ordered indices of *S* from smallest to largest
  $A^* = Y_k$
  $A^*$ = summed vector $A^*$, where $A^*_i = \frac{\sum_{i=1}^{j} A^*_i}{j}$
  $\Delta = \frac{\sum_{i=1}^{j} A^*_i - A_i}{j}$
  **if** Δ > 0
    $R = R^*$
    $A = A^*$
  **end if**
**end while**
**return** $R^*$

### Defining Threshold Behavior of Rules

   A rule, ***R***, denotes a hyper-rectangle with #***R*** dimensions. A feature can be used to split the data more than once along a path when the features are continuous. This presents a challenge when deciding which values to use as boundaries. Thresholding values near the root node will potentially be overly broad while the values at the deepest occurrences may be too narrow with respect to the given sample. In this implementation, a lower boundary for a feature in ***R*** is obtained by identifying the splitting values for the deepest occurrences of that feature across all trees when the sample's value was greater than the threshold (e.g. the set of splitting values for all third appearances of the feature in paths). The most distant of these values relative to the sample is selected as the lower boundary as a compromise between generality and precision. This same procedure is repeated in order to obtain an upper boundary using the feature's values when the sample's value was less than the threshold instead. If no such splits occurred in the forest then the respective maximum or minimum value of the data in that dimension is used to set the respective boundary. This process yields a coarse representation of the feature space that is being utilized by the random forest for making predictions.

### **Extending local rules to describe global patterns**

Any given *R* is a partial representation of globally important patterns in the dataset. To obtain a more complete picture of which patterns are most important, an ensemble of rules is created by deriving a single *R* from each observation with the class label of interest. Each *R*, along with its associated thresholding bounds, is then evaluated for precision and its general applicability within the original dataset. A final set of unique *Rs* is sorted by the strength of their associations with the class label of interest from least likely to most likely assuming a binomial distribution. It is assumed that the likelihood of a random sample having the class label is p(Label) = #labeled samples/#samples. The result is an interpretable table of relevant *Rs* and an estimated conditional probability that a sample, which meets the criteria of *R*, has the class label of interest.

## Code availability
The code to implement the JigSaw algorithms in R and to replicate the results presented here is available for download at: https://github.com/ddimucci/JigSaw.

## Results

### **Simulation experiments**

To evaluate JigSaw's ability to correctly identifying sets of interacting features where the ground truth is known, several simulation experiments were performed. For each experiment, 50 independent Gaussian features, x = ($x_1$, … $x_{50}$) with mean = 10 and standard deviation = 2, were generated for *n* = 1,000 independent observations. Interacting feature sets were also defined for each experiment such that if the conditions were met it would result in the assignment of the class label with a fixed probability. Each scenario was performed 10 times with a different random seed.

In all experiments, JigSaw was applied to each observation that received a label. Each observation yielded a local rule along with a corresponding set of decision boundaries. To extend the individual findings to global relevance, the observations in the training set to which each rule applies were identified (*m*). Then, the likelihood of obtaining the number of observed labeled observations in *m* Bernoulli trials with calculated. Rules were sorted into a ranked list from least likely to most likely and the combined ability of the rules to separate the classes when applied sequentially was quantified by measuring the area under the receiver operator curve (ROC-AUC) and the area under the precision-recall curve (PR-AUC). Redundant rules are removed from the list if their corresponding set of *m* is fully subsumed by a higher ranked rule. The lower bounds of the true decision space in each dimension were defined as the mean of the underlying distribution. The upper bounds of the true decision space were defined as the maximal value in the training set. For each interacting set, the average fraction of these boundaries that are covered by the bounds returned by JigSaw rules containing them were calculated. These results are summarized in Table 2 for each scenario.

To judge the algorithm's ability to recover meaningful sets of features, any rule that contained a complete set of interacting features was marked as a successful recovery while any rule that was missing one or more features from a set would be deemed a failure. This definition was chosen with the idea that in certain biological applications it is more costly to fail to identify a relevant feature than it is to measure an extraneous one. For each feature set, the rank of the first rule fully containing it was recorded and is summarized in Table 3. For the 10 simulations, the average rank of the first rule that fully contained a given interacting set is reported. Likewise, the average number of extraneous features in the corresponding rule is also reported. These rankings were determined without removing redundant rules from the list.

### Experiment 1
In this first scenario the necessary condition to assign the label was a single AND rule of four randomly selected features. Observations were assigned a label according to the condition:
$$p(\text{Label}) = 0.8 \cdot \mathbb{1}(x_1 > t_1 \ \& \ x_2 > t_2 \ \& \ x_3 > t_3 \ \& \ x_4 > t_4)$$
Where $\mathbb{1}$ is an indicator function and all values of $t$ are the mean of the underlying Gaussian. This thresholding condition resulted in approximately 5% of samples receiving the label.

### Experiment 2
The second simulation was set up identically to the first except that we injected noise by randomly assigning the label to **5%** of the samples that did not meet the threshold conditions.

### Experiment 3
The input data for the third simulation was generated identically to the previous simulations. However, in this scenario four mutually independent sets of features were defined. Each set contained four features chosen at random. Features were allowed to appear in multiple rules. Each set assigned the label with identical thresholding conditions as experiment 1.

### Experiment 4
Interacting feature sets were selected the same way as they were in experiment 3 but with **5%** noise injected into the labels in the same manner as experiment 2.

### Case Study 1
JigSaw was applied to the Statlog (Heart) data set available from the UCI Machine learning repository (22). This data set contains 270 records that are diagnosed as either having heart disease (120) or healthy (150). For each record in the data set, there are measurements for 13 attributes related to heart health (Table S1).

Before extracting rules from a forest trained on the full data set the generality of the JigSaw derived rules was first evaluated by 10-fold cross validation (where 30% of the input data was set aside as a test set for each fold). The boundaries of the rules were not tuned nor were they merged to maximize effectiveness on the training set and the simple procedure of applying rules sequentially was used. The mean ROC and PR AUCs for the JigSaw derived rule set and the full forest model is summarized in Table 4. JigSaw was then applied to a forest trained on the full data set and a subset of 27 rules that fully account for all disease samples in the training set was identified. Figure 1 shows cumulative amount of disease samples explained by rules extracted from the forest. Since the first three rules are the most general, the rest of the analysis is restricted to these.

Each of the three rules identified by JigSaw involve three attributes each. The attributes in each rule and their corresponding upper and lower bounds are summarized in Table 5. The first rule involved an interaction between sex, chest pain type (chest pain), and number of vessels colored by flourosopy (N-vessels). 46 records are subject to the boundaries identified by JigSaw (45 disease, 1 healthy). The second rule involved chest pain, N-vessels, and thallium-201 stress scintigraphy (thal). This rule applies to 52 records (48 disease, 4 healthy). The third rule involved chest pain, resting electrocardiographic results (ECG), and thal. This rule applied to 44 records (42 disease, 2 healthy). When all three of these rules were combined, they accounted for 86 records, 80 diseased and 6 healthy.

### Case Study 2
JigSaw was next applied to the breast cancer Coimbra data set (23) which is also available through the UCI machine learning repository. This data set contains 116 participant records and reports continuous measurements for 9 attributes: Age, body mass index (BMI), serum glucose, insulin, Homeostasis Model assessment of insulin resistance (HOMA), leptin, adiponectin, resistin, and Chemokine Monocyte Chemoattractant Protein 1 (MCP-1). The goal of the original study was to develop a diagnostic biomarker for distinguishing 64 patients with breast cancer (BC) from 52 patients without disease (healthy) using parameters that can easily be measured in the blood. With random forest, the authors achieved a high performing classifier but did not attempt to extract any logical rules from it. Conveniently, this data produces some relevant decision rules that involve just two features, which allows for easily viewing the decision boundaries identified by JigSaw.

Before extracting rules from the full data set the generality of the JigSaw derived rules was evaluated by 10-fold cross validation. Since this data set was smaller than the Heart data set a larger fraction of the data was held out as a separate test set (60%). The boundaries of the rules were not tuned nor were they merged to maximize effectiveness on the training set and the simple procedure of applying rules sequentially was used. The mean ROC and PR AUCs for the JigSaw derived rule set and the full forest model is summarized in Table 5. For each data split the effectiveness of the top ranked JigSaw rule was examined. On average the top rule applied to 9.1/35 samples in the test set (~26%) and correctly classified 6.7 of these samples as BC (~74%). For splits where the first rule happened to be 2-dimensional the inferred boundaries, training, and test set data points have been plotted. In at least one case (split 10) the boundaries could have been broadened to more generally apply to the training set (Figure S1).

Next, JigSaw was applied to a forest trained on the full data set and subset of 27 rules that fully account for all disease samples in the training set was identified. Figure 2 shows the cumulative amount of breast cancer samples recovered by rules obtained from applying JigSaw to a trained random forest model. The first rule involves 4 measured features (BMI, Leptin, Adiponectin, and Resistin) and applies to 27 records (24 BC, 3 healthy). The second rule uses just BMI and Resistin but has more conservative boundaries for these values than the first rule, it applies to 29 records (25 BC, 5 healthy). Figure 3 shows the decision boundaries of both rules in the BMI and Resistin dimensions. By combining their decision boundaries in order to form a new rule in two dimensions (BMI and Resistin), *Rule 1*, it can be extended to apply to 40 records (36 BC, 4 healthy).

To look for additional rules that could explain the remaining BC records, labels for records explained by *Rule 1* were flipped to reflect the healthy class, the resulting class distribution was 28 BC and 88 healthy samples. JigSaw was then applied to a random forest model trained on the same data set with the new manipulated labels. The best performing rule, *Rule 2*, involved Age and Resistin. It accounted for 28 samples (18 previously un-accounted for BC samples, 1 BC sample covered by *Rule 1*, and 9 new healthy samples). Figure 4 shows the shows the decision boundaries of *Rule 2*. Applying both rules to the data set serves to account for 67 samples (54 BC, 13 healthy). A third rule, *Rule 3*, involving Adiponectin and Resistin was identified that accounted for 9 total samples and 6 of the 10 remaining BC samples (5 new BC samples, 1 covered previously, and 3 healthy), bringing the total number of BC samples with an explanation to 59 BC with 16 healthy records within the same boundaries. All three rules and their respective decision boundaries are summarized in table 7.

## Discussion

JigSaw is a new algorithmic strategy for extracting classification rules from a random forest. By taking advantage of the structure of trees that correctly predict individual observations, it is capable of identifying sets of features and corresponding boundaries that are associated with a class of interest. Even in systems with spurious labeling and multiple feature sets associated with the target label, JigSaw is capable of reliably identifying at least one high-order interaction as a hyper-rectangle in feature space. Each hyper-rectangle is an explanation for why certain samples are predicted by the random forest to have a given class. In the case of the Heart data, the raw rules were nearly as effective as the full random forest model in classifying a held out test set. The Coimbra breast cancer data set proved to contain simpler, but less precise rules. In this case, with some simple manual curation of the JigSaw rules there was a single rule identified that by itself accounts for over half of the diseased samples with high precision.
The steps described in this work produce and rank explanatory rules for continuous numerical inputs. When applied to high-dimensional biological data sets, these rules could be used to generate new mechanistic hypotheses for follow up studies. Future work can be done in developing an algorithmic solution for the merging of overlapping rules into one and for automatically curating a minimal set of rules sufficient to explain the observed data.

glucose, age and BMI to predict the presence of breast cancer. BMC Cancer. 2018;

Tables

| Symbol | Meaning |
|---|---|
| # | Cardinality |
| A | Stores the values of Y sorted by their Euclidean distances in R to O |
| child | The node that immediately follows the parent |
| D | Dataset of continuous values |
| E | List of parent-child edges encountered by O across all trees in T. Ranked by frequency. |
| M | Trained random forest model |
| N | A p x p matrix that is initialized as a matrix of zeros. |
| O | The specified observation |
| out-of-bag | Trees in M for which O was not included in the training set. |
| p | Number of predictors in the data set |
| parent | When evaluating a pair of nodes along a path, the parent is the node occuring most closely to the root node |
| path | The sequence of feature indices encountered by O from the root node to the terminal node of a tree in T |
| R | Set of variables most relevant to O, initialized as an empty set |
| T | Set of out-of-bag trees with respect to O |
| V | Vector of the measured values for O |
| Y | Vector of observed labels for each observation in D |
| S | Vector of euclidean distances of each observation in D from O in the dimensions of R |
| k | Indices of all observtaions in D sorted in ascending order by their corresponding value in S. |
| Δ | Vector of differences between the current vector A and a new potential vector A* |
| * | Indicates that a value is temporary. |

Table 1. List of symbols used in algorithms 1 and 2.

| Scenario | | Mean Coverage | JigSaw Rule Sets | | Full Model | |
|---|---|---|---|---|---|---|
| # Sets | Noise | | ROC-AUC | PR-AUC | ROC-AUC | PR-AUC |
| 1 | 0 | 0.74 | .96 | 0.59 | 0.98 | 0.74 |
| 1 | 5% | 0.74 | 0.73 | 0.32 | 0.74 | 0.38 |
| 4 | 0 | 0.68 | 0.76 | 0.39 | 0.83 | 0.55 |
| 4 | 5% | .63 | 0.68 | 0.34 | 0.76 | 0.5 |

Table 2. Mean fraction of the true decision space identified for each ground truth rule and the classification metrics of predicting a separate test set in each simulated scenario.

| Scenario | | Rule Ranks | | | | Extra Features | | | | Mean # Sets Recovered |
|---|---|---|---|---|---|---|---|---|---|---|
| # Sets | Noise | Set 1 | Set 2 | Set 3 | Set 4 | Set 1 | Set 2 | Set 3 | Set 4 | |
| 1 | 0 | 1 | n/a | n/a | n/a | 0.3 | n/a | n/a | n/a | 4 |
| 1 | 5% | 1.1 | n/a | n/a | n/a | 0.6 | n/a | n/a | n/a | 4 |
| 4 | 0 | 1.8 | 7.8 | 25.1 | 36.3 | 0.9 | 1.2 | 1.6 | 3.1 | 4 |
| 4 | 5% | 2.5 | 7.0 | 28.6 | 62.2 | 1.5 | 1.7 | 2.5 | 2.7 | 3.9 |

Table 3. Average rank of the first rule to fully recover a feature set and superfluous features for 10 simulations in each scenario.

| JigSaw Rule Sets | | Full Model | |
|---|---|---|---|
| ROC-AUC | PR-AUC | ROC-AUC | PR-AUC |
| 0.849 | 0.775 | 0.899 | 0.888 |

Table 4. Average test set ROC and PR AUC from 10-fold cross validation on Heart data set holding out 30% of the data for each test set.

| | Attributes | | |
|---|---|---|---|
| Rule 1 | Sex | Chest Pain | N Vessels |
| Lower Bounds | 0.5 | 3.5 | 0.5 |
| Upper Bounds | 1 | 4 | 2.5 |
| Rule 2 | Chest Pain | N Vessels | Thal |
| Lower Bounds | 1.5 | 0.5 | 4.5 |
| Upper Bounds | 4 | 2.5 | 7 |
| Rule 3 | Chest Pain | ECG | Thal |
| Lower Bounds | 3.5 | 0.5 | 4.5 |
| Upper Bounds | 4 | 2 | 7 |

Table 5. Boundary conditions for the three most significant rules from the heart disease data set.

| JigSaw Rule Sets | | Full Model | |
|---|---|---|---|
| ROC-AUC | PR-AUC | ROC-AUC | PR-AUC |
| 0.68 | 0.70 | 0.80 | 0.83 |

Table 6 Average test set ROC and PR AUC from 10-fold cross validation on the Coimbra data set holding out 30% of the data for each test set.

| Rule 1 | BMI | Resistin |
|---|---|---|
| Lower Bounds | 18.9 | 13.25 |
| Upper Bounds | 32.07 | 68.66 |
| Rule 2 | Age | Resistin |
| Lower Bounds | 41 | 5.03 |
| Upper Bounds | 63 | 13.65 |
| Rule 3 | Adiponectin | Resistin |
| Lower Bounds | 11.84 | 2 |
| Upper Bounds | 35.90 | 6.80 |

Table 7. Rules identified in the breast cancer data with lower and upper bounds.

| Feature Index | Attribute | Abbreviation Used |
|---|---|---|
| 1 | Age | Age |
| 2 | Sex | Sex |
| 3 | Chest pain type | Chest pain |
| 4 | resting blood pressure | n/a |
| 5 | serum cholesterol in mg/dl | n/a |
| 6 | fasting blood sugar | n/a |
| 7 | resting electrocardiographic results | ECG |
| 8 | maximum heart rate achieved | n/a |
| 9 | exercised induced angina | n/a |
| 10 | oldpeak = ST depression induced by exercise relative to rest | n/a |
| 11 | the slope of the peak exercise ST segment | n/a |
| 12 | number of major vessels (0-3) colored by flourosopy | N vessels |
| 13 | thal: 3 = normal; 6 = fixed defect; 7 = reversable defect | Thal |

Table S1. Attributes measured in heart disease data set.

Figures

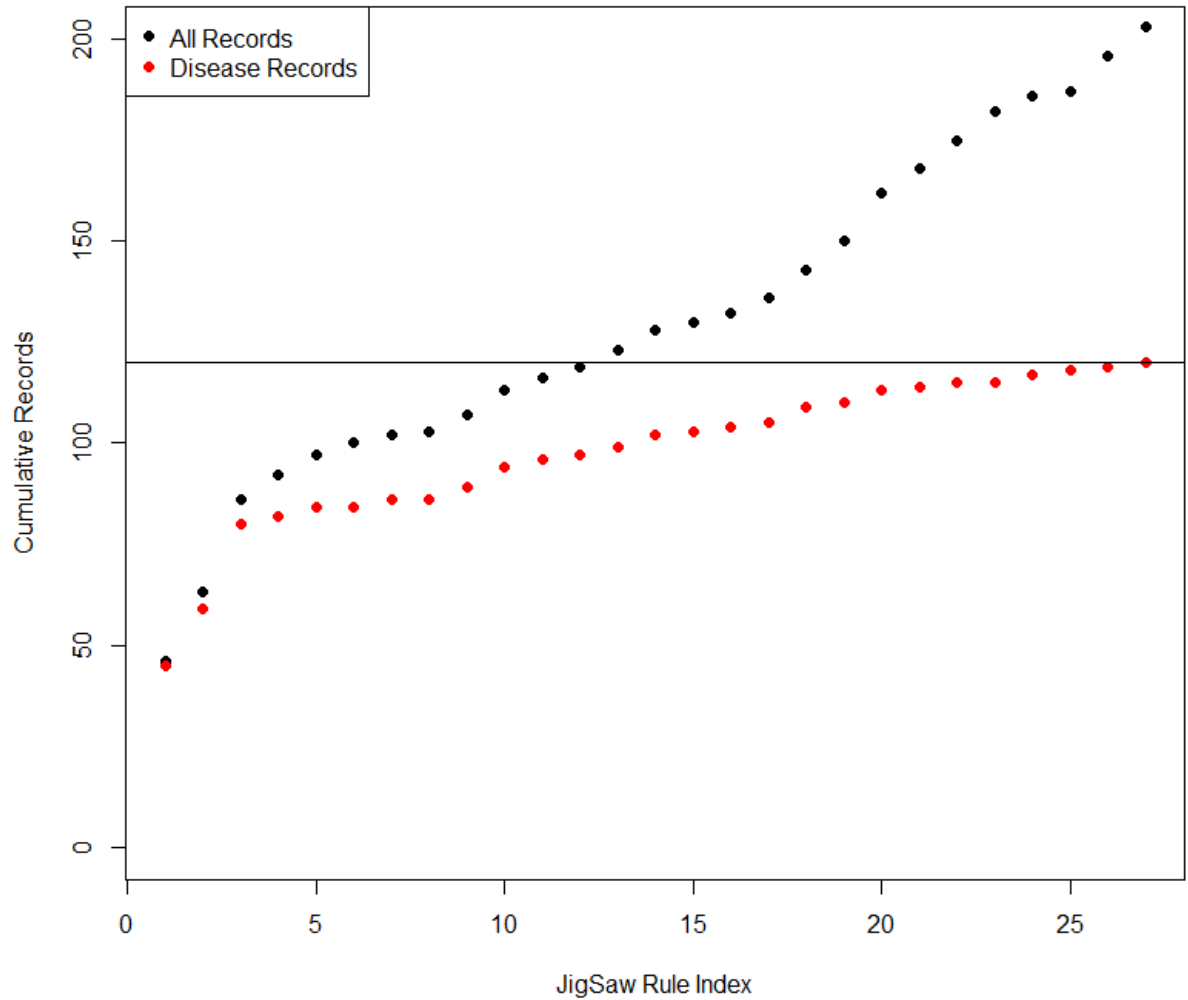

**Figure 1. Cumulative number of heart disease records explained.** Red points indicate the cumulative number of disease records explained by rules identified by JigSaw. Black points indicate the cumulative total of all records that meet the criteria defined by each rule. The solid line marks the total number of disease records in the data set.

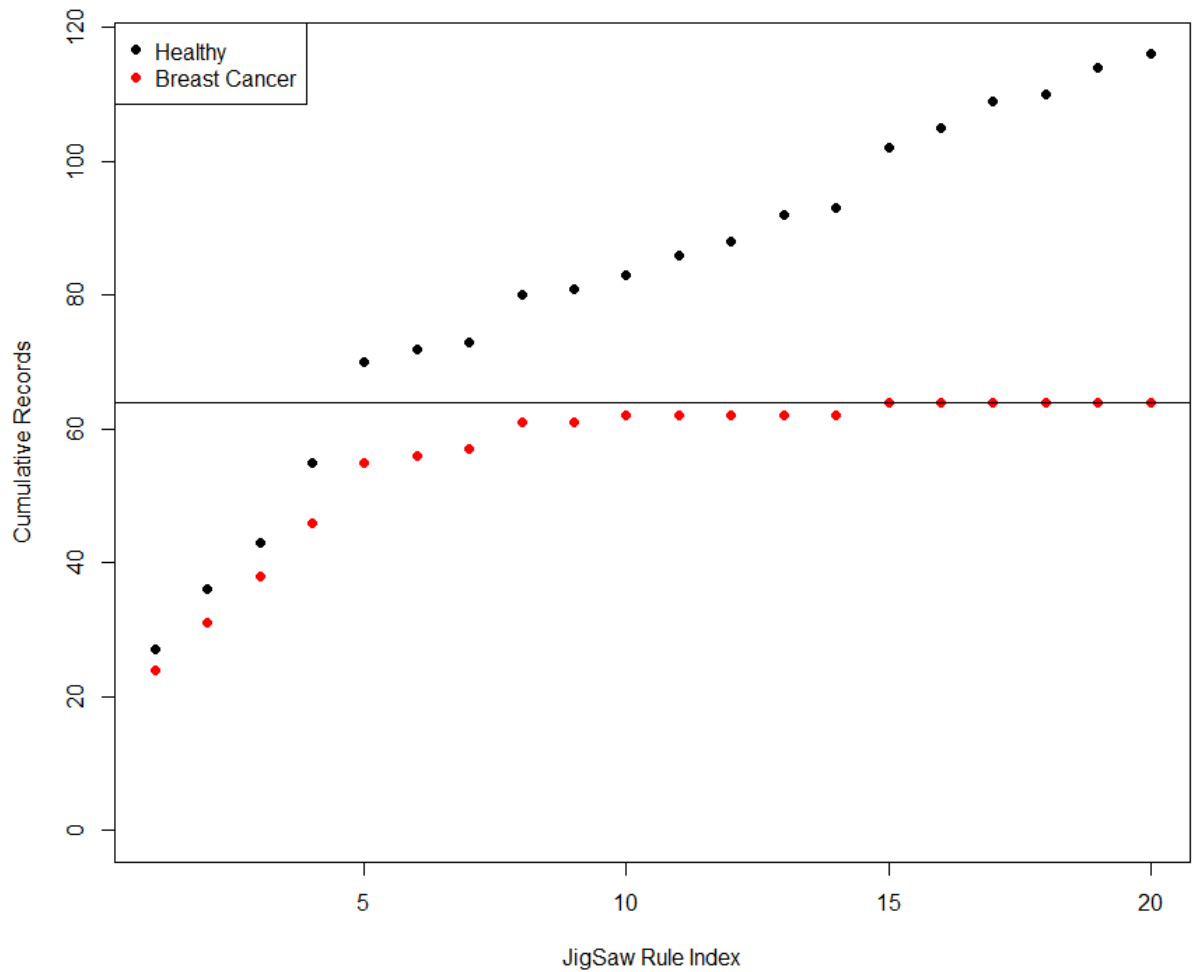

**Figure 2. Cumulative number of breast cancer records explained.** Red points indicate the cumulative number of breast cancer records explained by rules identified by JigSaw. Black points indicate the cumulative total of all records that meet the criteria defined by each rule. The solid line marks the total number of breast cancer records in the data set.

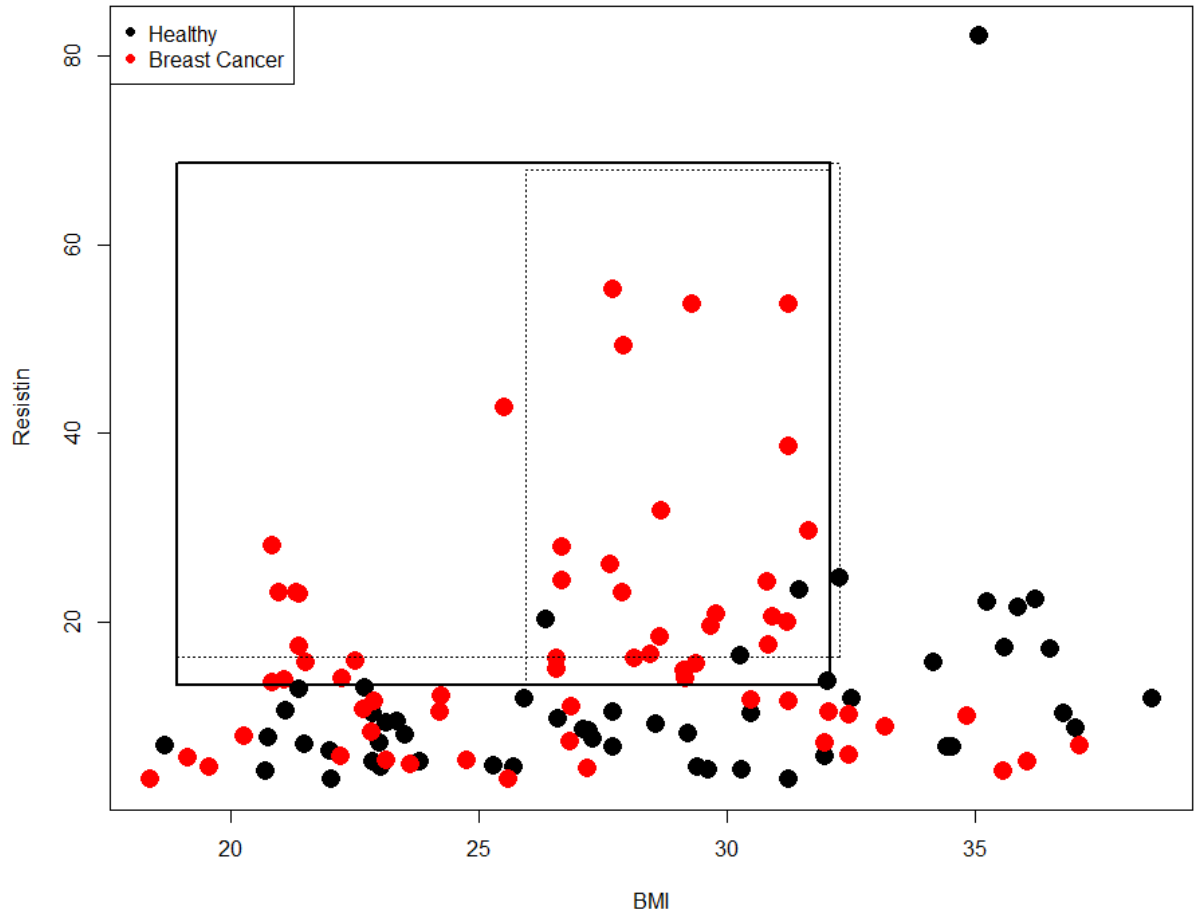

**Figure 3. Decision bounds of *Rule 1*.** The first two rules identified by JigSaw involve measurements in BMI and Resistin and share overlapping boundaries in these dimensions. Dashed lines indicate the decision bounds of those rules. Solid box are the decision bounds for *Rule 1* obtained by inspecting and then merging the boundaries of the top two JigSaw rules.

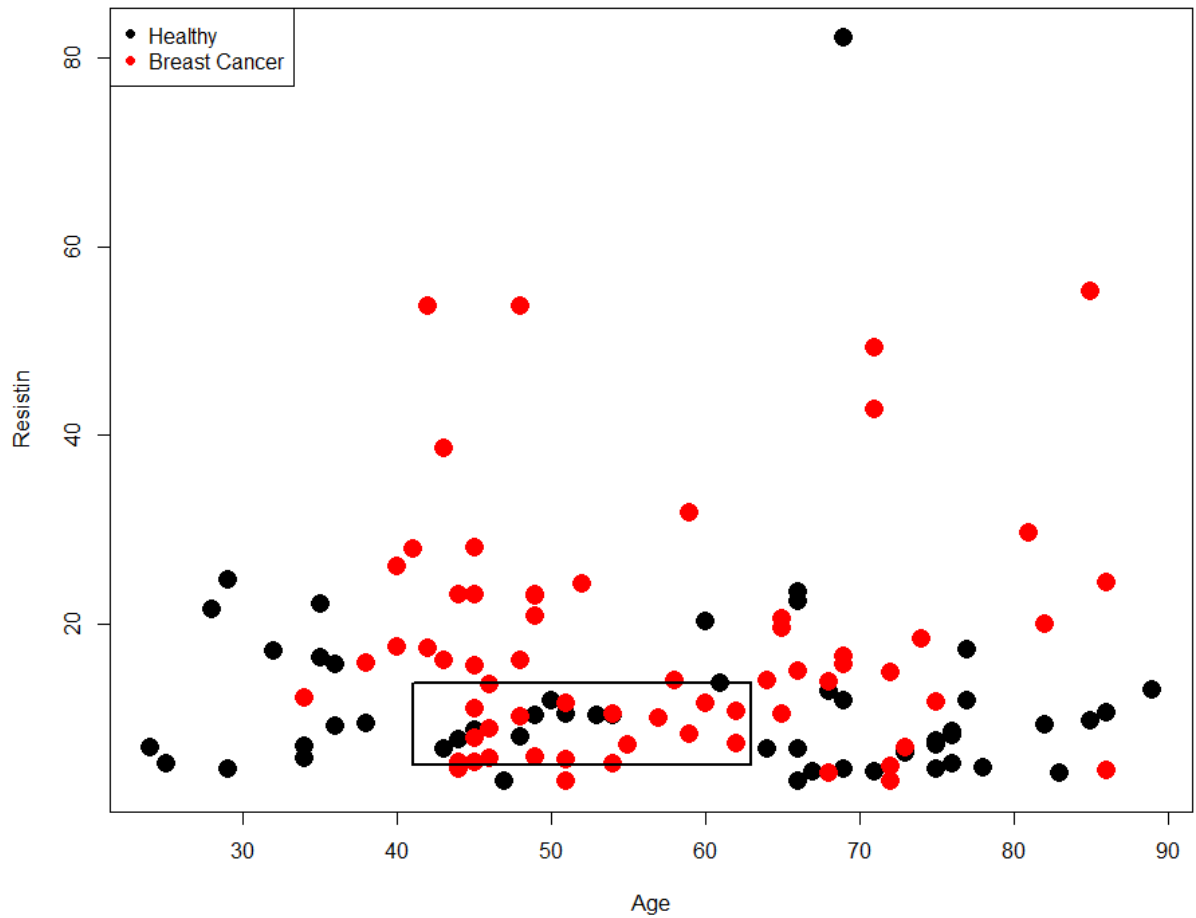

**Figure 4. Decision bounds of *Rule 2*.** The second rule identified by JigSaw is shown, it considers measurements of just Age and Resistin. The boundaries shown here are the original bounds defined by JigSaw but can be extended by removing the upper bounds in the Resistin direction. Points that would be captured by this action are already accounted for by *Rule 1*.

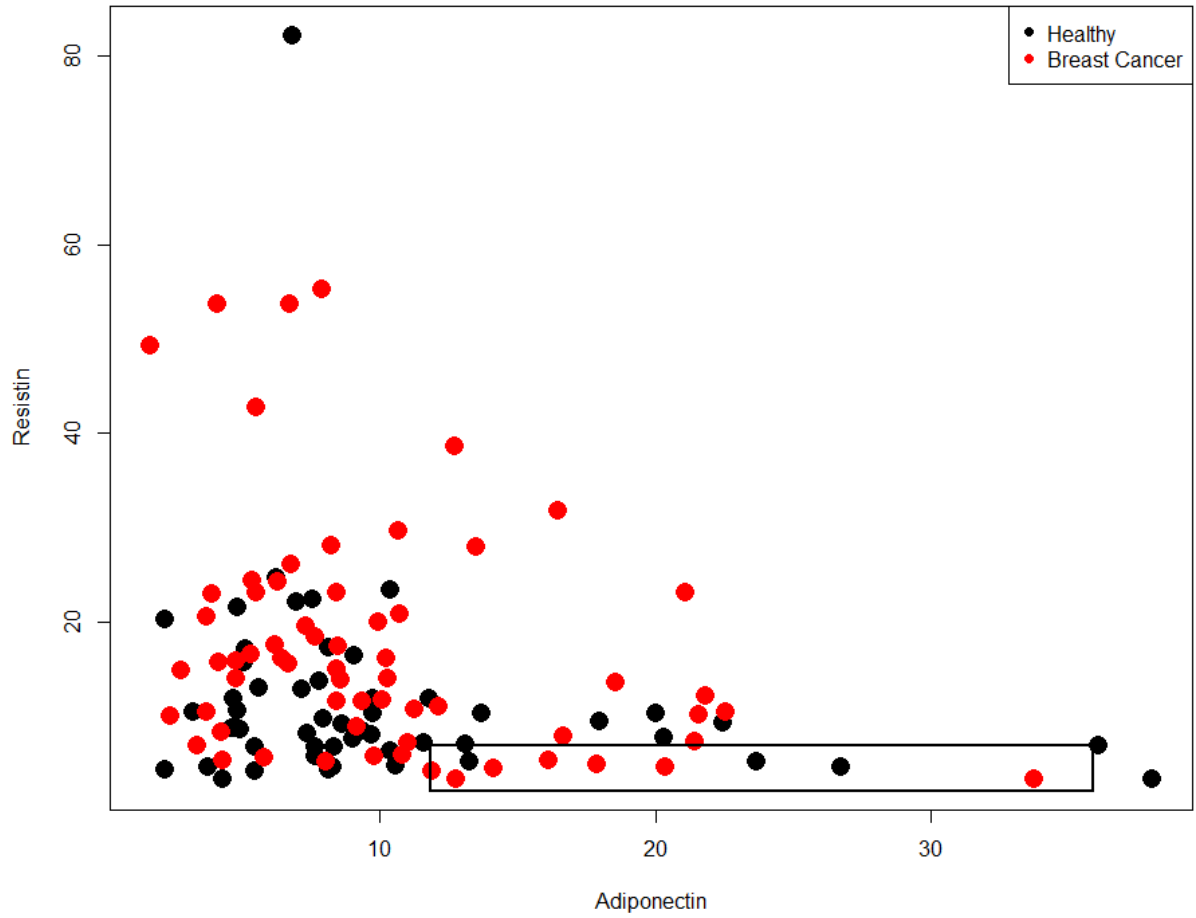

**Figure 5. Decision bounds of *Rule 3*.** The third rule identified by JigSaw is shown, it considers measurements of just Adiponectin and Resistin. The boundaries shown here are the original bounds defined by JigSaw.